\documentstyle[preprint,eqsecnum,aps]{revtex}

\def\beq{\begin{eqnarray}}
\def\eeq{\end{eqnarray}}
\newcommand{\nn}{\nonumber}
\def\bea{\begin{array}}
\def\eea{\end{array}}
\def\dj{\left(\frac 1 {2x} \frac d {dx} \right)^j}
\def\bn{{d+n-2 \choose n}}
\def\anu{\alpha_\nu}

\begin{document}
\title{Chiral Bag Boundary Conditions on the Ball}
\author{Giampiero Esposito,$^{1,2}$
\thanks{Electronic address: giampiero.esposito@na.infn.it}
and Klaus Kirsten$^{3}$
\thanks{Electronic address: kirsten@mis.mpg.de}}
\address{${ }^{1}$Istituto Nazionale di
Fisica Nucleare, Sezione di Napoli,\\
Complesso Universitario di Monte S. Angelo, Via Cintia,
Edificio N', 80126 Napoli, Italy\\
${ }^{2}$Dipartimento di Scienze Fisiche,
Complesso Universitario di Monte S. Angelo,\\
Via Cintia, Edificio N', 80126 Napoli, Italy\\
${ }^{3}$Max Planck Institute for Mathematics in the Sciences,\\
Inselstrasse 22-26, 04103 Leipzig, Germany}
\maketitle
\begin{abstract}
Local boundary conditions for spinor fields are expressed in terms
of a 1-parameter family of boundary operators, and find applications
ranging from (supersymmetric) quantum cosmology to the bag model in
quantum chromodynamics. The present paper proves that, for massless
spinor fields
on the Euclidean ball in dimensions $d=2,4,6$, the resulting
$\zeta(0)$ value is independent of such a $\theta$
parameter, while the various heat-kernel coefficients exhibit a
$\theta$-dependence which is eventually expressed in a simple way
through hyperbolic functions and their integer powers.
\end{abstract}

\section{Introduction}

The choice of boundary conditions in the theories of fundamental
interactions has always attracted the interest of theoretical
physicists, not only as a part of the general programme aimed at
deriving the basic equations of physics from a few guiding
principles [1--8], but also as a tool for studying concrete
problems in quantum field theory and global analysis [9--11].

In particular, we are here interested in studying local boundary
conditions for massless spin-${1\over 2}$ fields, whose main
motivations may be summarized as follows [12,13].
\vskip 0.3cm
\noindent
(i) The Breitenlohner--Freedman--Hawking [14,15] boundary conditions
for gauged supergravity theories in anti-de Sitter space are local
and are expressed, for spin-${1\over 2}$ fields, in terms of a
projection operator. The rigid supersymmetry transformations
between massless linearized fields of different spins map classical
solutions of the linearized field equations, subject to such
boundary conditions at infinity, to classical solutions for an
adjacent spin, subject to the same family of boundary conditions
at infinity [12].
\vskip 0.3cm
\noindent
(ii) In simple supergravity the spatial tetrad and a projection
formed from the spatial components of the spin-${3\over 2}$
potential transform into each other under half of the local
supersymmetry transformations at the boundary [16]. The supergravity
action functional can also be made invariant under this class
of local supersymmetry transformations. On considering the
extension to supergravity models based on the group $O(N)$,
the supersymmetry transformation laws show that,
{\it for spin-${1\over 2}$ fields only}, the same projector
should be specified on the boundary as in the
Breitenlohner--Freedman--Hawking case.
\vskip 0.3cm
\noindent
(iii) The work in Ref. [13] has shown that, instead of quantizing
gauge theories on a sphere or on a torus one can quantize them
in an even-dimensional Euclidean bag and impose
$SU_{A}(N_{f})$-breaking boundary conditions to trigger a chiral
symmetry breaking. On investigating how the various correlators
depend on the parameter $\theta$ characterizing the boundary
conditions one then finds that bag boundary conditions are a
substitute for small quark masses [13].

More precisely, in theories of Euclidean bags, chiral symmetry
breaking is triggered by imposing the boundary conditions
\cite{[13],hras84-245-118}
\begin{equation}
0=\pi_{-}\psi |_{\partial M}
={1\over 2}\left(1+i e^{\theta \gamma^{5}}\gamma^{5}
\gamma^{m} \right) \psi |_{\partial M}
\label{(1.1)}
\end{equation}
on the spinor field $\psi$. Here we focus on the $d$-dimensional
Euclidean ball, which is the portion of flat $d$-dimensional
Euclidean space bounded by the $S^{d-1}$ sphere. The eigenspinors
of the Dirac operator on the ball have the form \cite{[18]}
\begin{equation}
\psi_{\pm}^{(+)}={C\over r^{(d-2)/2}}
\pmatrix{i J_{n+d/2}(kr)Z_{+}^{(n)}(\Omega) \cr
\varepsilon J_{n+(d-2)/2}(kr)Z_{+}^{(n)}(\Omega) \cr},
\label{(1.2)}
\end{equation}
\begin{equation}
\psi_{\pm}^{(-)}={C\over r^{(d-2)/2}}
\pmatrix{\varepsilon J_{n+(d-2)/2}(kr)Z_{-}^{(n)}(\Omega) \cr
i J_{n+d/2}(kr)Z_{-}^{(n)}(\Omega) \cr},
\label{(1.3)}
\end{equation}
where $C$ is a normalization constant, $\varepsilon \equiv \pm 1$,
$n=0,1,2,...,\infty$, and $Z_{\pm}^{(n)}(\Omega)$ are the spinor
modes on the sphere \cite{[19]}. In Eq. (1.1), the boundary operator
reduces to the matrix
$$
{1 \over 2} \pmatrix{1 & -i e^{\theta} \cr
i e^{-\theta} & 1 \cr},
$$
and its application to (1.2) and (1.3) yields the eigenvalue
condition \cite{[17]}
\begin{equation}
J_{n+d/2}(k)-\varepsilon e^{\theta}J_{n+d/2-1}(k)=0
\label{(1.4)}
\end{equation}
for $\psi_{\pm}^{(+)}$, and
\begin{equation}
J_{n+d/2}(k)+\varepsilon e^{-\theta}J_{n+d/2-1}(k)=0
\label{(1.5)}
\end{equation}
for $\psi_{\pm}^{(-)}$, where $r$ has been set to $1$ for
convenience [3]. By eigenvalue condition we mean the equation
obeyed by the eigenvalues by virtue of the boundary conditions,
which yields them only implicitly [3]. Equations (1.4) and (1.5) lead
eventually to the eigenvalue condition in non-linear form, i.e.
\begin{equation}
J_{n+d/2-1}^{2}(k)-e^{-2\theta}J_{n+d/2}^{2}(k)=0,
\label{(1.6)}
\end{equation}
\begin{equation}
J_{n+d/2-1}^{2}(k)-e^{2\theta}J_{n+d/2}^{2}(k)=0.
\label{(1.7)}
\end{equation}
Of course, it is enough to deal with one of these equations,
while the contributions from the other
follow by replacing $\theta$ with $-\theta$.

Recently, in $d=2$ dimensions, the spectral asymmetry following
from the boundary conditions (1.1) was considered in \cite{mariel}.
Asymmetry properties are encoded in the eta function which was analyzed
using contour integral methods, see e.g. \cite{[11]}.

Instead,
we study heat-kernel asymptotics for the squared Dirac operator on
the $d$-ball with eigenvalue conditions (1.6) and (1.7) which is
related to an analysis of the zeta function. Strictly,
one can actually obtain two second-order operators of Laplace
type out of the Euclidean Dirac operator $D$, i.e.
$$
P_{1} \equiv DD^{\dagger} \; \; {\rm and} \; \;
P_{2} \equiv D^{\dagger}D,
$$
where $D^{\dagger}$ denotes the (formal) adjoint of $D$. The
existence of both $P_{1}$ and $P_{2}$ is crucial for index theory [5],
and by taking into account both (1.6) and (1.7) we correctly
take care of this (see Ref. [12] for the mode-by-mode version of
$P_{1}$ and $P_{2}$ on the 4-ball). To be self-contained,
recall that, given the second-order elliptic operator $P$, the
heat kernel can be defined as the solution, for $\tau>0$, of the
associated heat equation
\begin{equation}
\left({\partial \over \partial \tau}+P \right)U(x,y;\tau)=0,
\label{(1.8)}
\end{equation}
subject to the initial condition ($(M,g)$ being the background
geometry)
\begin{equation}
\lim_{\tau \to 0}\int_{M} U(x,y;\tau)\varphi(y)
\sqrt{{\rm det} \; g}\; dy=\varphi(x),
\label{(1.9)}
\end{equation}
and to suitable boundary conditions
\begin{equation}
[{\cal B}U(x,y;\tau)]_{\partial M}=0,
\label{(1.10)}
\end{equation}
which preserve ellipticity and lead to self-adjointness of the
boundary-value problem [9--11].
The functional (or $L^{2}$) trace of the
heat kernel is obtained by considering the heat-kernel diagonal
$U(x,x;\tau)$, taking its fibre trace ${\rm Tr}_{V}U(x,x;\tau)$
(since $U(x,y;\tau)$ carries (implicit) group indices in the case
of gauge theories), and integrating such a
fibre trace over $M$, i.e.
\begin{equation}
{\rm Tr}_{L^{2}}e^{-\tau P}=\int_{M}{\rm Tr}_{V}U(x,x;\tau)
\sqrt{{\rm det} \; g} \; dx.
\label{(1.11)}
\end{equation}
The asymptotic expansion we are interested in holds for
$\tau \rightarrow 0^{+}$ and has the form [9]
\begin{equation}
{\rm Tr}_{L^{2}}(e^{-\tau P}) \sim \tau^{-d/2}
\sum_{n=0}^{\infty}\tau^{n/2}a_{n/2}(P,{\cal B}),
\label{(1.12)}
\end{equation}
where the heat-kernel coefficients $a_{n/2}(P,{\cal B})$ are
said to describe the global (integrated) asymptotics and consist
of an interior part $c_{n/2}(P)$ and a boundary part
$b_{n/2}(P,{\cal B})$, i.e.
\begin{equation}
a_{n/2}(P,{\cal B})=c_{n/2}(P)+b_{n/2}(P,{\cal B}).
\label{(1.13)}
\end{equation}
At a deeper level, we might introduce a smearing function and
consider instead the $L^{2}$ trace of $fe^{-\tau P}$, with $f$
a smooth function on $M$. This takes into account the distributional
behaviour of the heat kernel from the point of view of invariance
theory (here ``invariance'' refers to the invariants of the
orthogonal group, which determine completely the functional form
of $a_{n/2}(P,{\cal B})$ [9]). However, mode-by-mode calculations
like the ones we are going to consider can be performed without
exploiting the introduction of $f$, and hence we limit ourselves
to using Eqs. (1.11)--(1.13). Section II describes the $\zeta$-function
algorithm in Ref. \cite{[20]} on the Euclidean $d$-ball \cite{dowk96-13-585},
and Sec. III
generalizes the work in Ref. [12] by showing that, on the 4-ball,
non-vanishing values of $\theta$ in Eqs. (1.6) and (1.7) do not
affect the conformal anomaly. The hardest part of our
analysis is then presented in Secs. IV and V, where heat-kernel
coefficients are studied for arbitrary dimension $d$, with several
explicit formulae in $d=2,4,6$. Concluding remarks and open problems
are described in Sec. VI, while relevant details can be found in
the Appendix.

\section{The Moss algorithm for the $d$-ball}

The starting point in our investigation of the eigenvalue
condition (1.7) for the purpose of heat-kernel asymptotics
is the use of the $\zeta$-function at large $x$, which was
first described in Ref. {\cite{[20]} with application to 4-dimensional
background geometries. However, since we are interested in
the Euclidean $d$-ball, we put no restriction on the dimension
of $M$, denoted by $d$ as in Sec. I, and we follow the general procedure
as outlined by Dowker \cite{dowk96-13-585}. First
we point out that on
replacing the eigenvalues $\lambda_{n}$ of $P$ by
$\lambda_{n}+x^{2}$ ($x$ being a large real parameter), one has
the $\zeta$-function at large $x$ in the form
\begin{equation}
\zeta(s,x^{2}) \equiv \sum_{n}(\lambda_{n}+x^{2})^{-s}
={1\over \Gamma(s)}\int_{0}^{\infty}\tau^{s-1}U_{x}(\tau)d\tau,
\label{(2.1)}
\end{equation}
having defined the integrated heat kernel (or functional trace
of the heat kernel at large $x$) as
\begin{equation}
U_{x}(\tau) \equiv \sum_{n}e^{-(\lambda_{n}+x^{2})\tau}
=e^{-x^{2}\tau}U(\tau).
\label{(2.2)}
\end{equation}
By virtue of the asymptotic expansion already encountered in
the Introduction, i.e.
\begin{equation}
U(\tau) \equiv \sum_{n}e^{-\lambda_{n}\tau} \sim
\sum_{n=0}^{\infty}a_{n/ 2}\tau^{(n-d)\over 2} \;
{\rm as} \; \tau \rightarrow 0^{+},
\label{(2.3)}
\end{equation}
we therefore find
\begin{equation}
\zeta(s,x^{2}) \sim {1\over \Gamma(s)}
\sum_{n=0}^{\infty}a_{n/ 2}I(x;s,n,d),
\label{(2.4)}
\end{equation}
having defined
\begin{equation}
I(x;s,n,d) \equiv \int_{0}^{\infty}\tau^{s-1+{(n-d)\over 2}}
e^{-x^{2}\tau} \; d\tau.
\label{(2.5)}
\end{equation}
Now we distinguish two cases, depending on whether $d$ is even
or odd. In the former, we consider $s={\overline s}$ such that
${\overline s}-1-{d\over 2}=0$, i.e. ${\overline s} \equiv
1+{d\over 2}$ which implies (on defining $\tau x^{2} \equiv z$)
\begin{equation}
I(x;{\overline s},n,d)=\int_{0}^{\infty}
\tau^{{n\over 2}}e^{-x^{2}\tau}
\; d\tau = x^{-n-2}\Gamma \left(1+{n\over 2}\right),
\label{(2.6)}
\end{equation}
and hence yields, for $d=2k, k=0,1,2,...$
\begin{equation}
\zeta \left(1+{d\over 2},x^{2}\right) \sim
\sum_{n=0}^{\infty}a_{n/ 2}{\Gamma \left(1+{n\over 2}\right)
\over \Gamma \left(1+{d\over 2}\right)}x^{-n-2}.
\label{(2.7)}
\end{equation}
In the latter, we consider $s={\widetilde s}$ such that
$$
s=1+{(d-1)\over 2} \equiv {\widetilde s},
$$
which implies
\begin{equation}
I(x;{\widetilde s},n,d)=\int_{0}^{\infty}\tau^{{n-1\over 2}}
e^{-x^{2}\tau} \; d\tau
=x^{-n-1}\Gamma \left({1+n \over 2}\right).
\label{(2.8)}
\end{equation}

On the other hand, since the function expressing the eigenvalue
condition (1.7) admits a canonical product representation
(see Appendix), one
can prove, on setting $\nu \equiv n+{d\over 2}$ for $d$ even, the
identity
\begin{eqnarray}
\; & \; &
\Gamma \left(1+{d\over 2}\right)
\zeta \left(1+{d\over 2},x^{2} \right)
=(-1)^{d\over 2}\sum_{n=0}^{\infty}2^{{d\over 2}-1}
\pmatrix{d+n-2 \cr n \cr}
{\left({1\over 2x}{d\over dx}\right)}^{1+{d\over 2}}
\nonumber \\
& \times & \log \Bigr[(ix)^{-2(\nu-1)}(J_{\nu-1}^{2}(ix)
-e^{2 \theta}J_{\nu}^{2}(ix))\Bigr],
\label{(2.9)}
\end{eqnarray}
where $2^{d\over 2}$ is the dimension $d_{s}$ of spinor space,
and $\mbox{deg}(n)={1\over 2}d_{s}\pmatrix{d+n-2 \cr n \cr}$ is the degeneracy
associated with the implicit eigenvalue condition (1.7).
Thus, the heat-kernel coefficient $a_{l\over 2}$ is equal to
${1\over \Gamma \left(1+{l\over 2}\right)}$ (respectively
${1\over \Gamma \left({1+l \over 2}\right)}$) times the
coefficient of $x^{-l-2}$ (respectively $x^{-l-1}$) in the
asymptotic expansion of the right-hand side of (2.9) in
even (respectively odd) dimension. On focusing for definiteness
on the even $d$ case, we now exploit the identity
\begin{equation}
J_{\nu-1}(k)=J_{\nu}'(k)+{\nu \over k}J_{\nu}(k),
\label{(2.10)}
\end{equation}
and obtain
\begin{equation}
J_{\nu-1}^{2}(ix)-e^{2 \theta}J_{\nu}^{2}(ix)
= {J_{\nu}'}^{2}(ix)-\left({\nu^{2}\over x^{2}}+e^{2\theta}\right)
J_{\nu}^{2}(ix)+{2\nu \over ix}J_{\nu}(ix)J_{\nu}'(ix).
\label{(2.11)}
\end{equation}
Thus, on defining $\alpha_{\nu} \equiv \sqrt{\nu^{2}+x^{2}}$ and
using the uniform asymptotic expansions of $J_{\nu}(ix)$ and
$J_{\nu}'(ix)$ summarized in the Appendix we find
\begin{equation}
J_{\nu-1}^{2}(ix)-e^{2\theta}J_{\nu}^{2}(ix)
\sim {(ix)^{2(\nu-1)}\over 2\pi}\alpha_{\nu}e^{2\alpha_{\nu}}
e^{-2\nu \log(\nu+\alpha_{\nu})}
\Bigr[\Sigma_{1}^{2}A_{\theta}(t)+\Sigma_{2}^{2}
+2t \Sigma_{1}\Sigma_{2}\Bigr],
\label{(2.12)}
\end{equation}
where we have defined
\begin{equation}
t \equiv {\nu \over \alpha_{\nu}},
\label{(2.13)}
\end{equation}
\begin{equation}
A_{\theta}(t) \equiv 1+(t^{2}-1)(1-e^{2\theta}).
\label{(2.14)}
\end{equation}
As expected, our formulae reduce, at $\theta=0$, to the asymptotic
expansions used in Ref. [12]. From now on we need to recall that
the functions $\Sigma_{1}$ and $\Sigma_{2}$ have asymptotic
series in the form
\begin{equation}
\Sigma_{1} \sim \sum_{k=0}^{\infty}{u_{k}(t)\over \nu^{k}},
\label{(2.15)}
\end{equation}
\begin{equation}
\Sigma_{2} \sim \sum_{k=0}^{\infty}{v_{k}(t)\over \nu^{k}},
\label{(2.16)}
\end{equation}
where $u_{k}$ and $v_{k}$ are the Debye polynomials given in
Ref. \cite{[21]}. The asymptotic expansions on the right-hand sides of
(2.15) and (2.16) can be re-expressed as
\begin{equation}
\sum_{k=0}^{\infty}{u_{k}(t)\over \nu^{k}} \sim
\sum_{j=0}^{\infty}{a_{j}(t)\over (\alpha_{\nu})^{j}},
\label{(2.17)}
\end{equation}
\begin{equation}
\sum_{k=0}^{\infty}{v_{k}(t)\over \nu^{k}} \sim
\sum_{j=0}^{\infty}{b_{j}(t)\over (\alpha_{\nu})^{j}},
\label{(2.18)}
\end{equation}
where
\begin{equation}
a_{i}(t)={u_{i}(t)\over t^{i}}, \;
b_{i}(t)={v_{i}(t)\over t^{i}}, \; \forall i \geq 0.
\label{(2.19)}
\end{equation}
Now the asymptotic expansion (2.12) suggests defining
\begin{equation}
{\widetilde \Sigma} \equiv \Sigma_{1}^{2}A_{\theta}(t)
+\Sigma_{2}^{2}+2t \Sigma_{1}\Sigma_{2},
\label{(2.20)}
\end{equation}
and hence studying the asymptotic expansion of
$\log({\widetilde \Sigma})$ in the relation to be used in
(2.9), i.e.
\begin{eqnarray}
\; & \; & \log \Bigr[(ix)^{-2(\nu-1)}
(J_{\nu-1}^{2}-e^{2\theta}J_{\nu}^{2})(ix)\Bigr] \nonumber \\
& \sim & -\log(2\pi)+\log \alpha_{\nu}+2\alpha_{\nu}
-2\nu \log(\nu+\alpha_{\nu})+\log {\widetilde \Sigma}.
\label{(2.21)}
\end{eqnarray}
>From the relations (2.13)--(2.20) $\widetilde \Sigma$ has the
asymptotic expansion
\begin{equation}
{\widetilde \Sigma} \sim \sum_{p=0}^{\infty}
{c_{p}\over (\alpha_{\nu})^{p}},
\label{(2.22)}
\end{equation}
where the first few $c_{p}$ coefficients read
\begin{equation}
c_{0}=A_{\theta}+1+2t,
\label{(2.23)}
\end{equation}
\begin{equation}
c_{1}=2a_{1}A_{\theta}+2b_{1}+2t(a_{1}+b_{1}),
\label{(2.24)}
\end{equation}
\begin{equation}
c_{2}=(2a_{2}+a_{1}^{2})A_{\theta}+(2b_{2}+b_{1}^{2})
+2t(a_{2}+b_{2}+a_{1}b_{1}),
\label{(2.25)}
\end{equation}
\begin{equation}
c_{3}=2(a_{3}+a_{1}a_{2})A_{\theta}+2(b_{3}+b_{1}b_{2})
+2t(a_{3}+b_{3}+a_{1}b_{2}+a_{2}b_{1}).
\label{(2.26)}
\end{equation}
Now defining
\begin{equation}
\Sigma \equiv {{\widetilde \Sigma}\over c_{0}},
\label{(2.27)}
\end{equation}
and making the usual expansion
\begin{equation}
\log(1+f)=\sum_{j=0}^{\infty}(-1)^{j+1}{f^{j}\over j},
\label{(2.28)}
\end{equation}
valid as $f \rightarrow 0$, we find
\begin{equation}
\log {\widetilde \Sigma}=\log c_{0}+\log \Sigma
\sim \log c_{0}+\sum_{p=1}^{\infty}
{A_{p}\over (\alpha_{\nu})^{p}},
\label{(2.29)}
\end{equation}
where explicit formulae can be given for all $A_{p}$.
In particular
\begin{equation}
A_{1}={c_{1}\over c_{0}},
\label{(2.30)}
\end{equation}
\begin{equation}
A_{2}={c_{2}\over c_{0}}-{1\over 2}(A_{1})^{2},
\label{(2.31)}
\end{equation}
\begin{equation}
A_{3}={c_{3}\over c_{0}}-A_{1}A_{2}-{1\over 6}(A_{1})^{3}.
\label{(2.32)}
\end{equation}
Using the definition
\begin{equation}
f_{\theta}(t) \equiv 1+{(t-1)\over 2}(1-e^{2\theta}),
\label{(2.33)}
\end{equation}
jointly with our previous formulae and the explicit form of
Debye polynomials \cite{[21]}, a lengthy calculation yields
as many $A_{p}$ terms as are needed. For example (cf. Ref. [12])
\begin{equation}
A_{1}={1\over 4}-{5\over 12}t^{2}
+{{1\over 2}(t^{2}-1)\over f_{\theta}(t)},
\label{(2.34)}
\end{equation}
\begin{equation}
A_{2}=\sum_{k=0}^{2}{\Omega_{k}(t)\over (f_{\theta}(t))^{k}},
\label{(2.35)}
\end{equation}
\begin{equation}
A_{3}=\sum_{k=0}^{3}{\omega_{k}(t)\over (f_{\theta}(t))^{k}},
\label{(2.36)}
\end{equation}
where
\begin{equation}
\Omega_{0}(t)={1\over 8}-{3\over 4}t^{2}+{5\over 8}t^{4},
\label{(2.37)}
\end{equation}
\begin{equation}
\Omega_{1}(t)=-{t\over 8}+{5\over 8}t^{2}+{t^{3}\over 8}
-{5\over 8}t^{4},
\label{(2.38)}
\end{equation}
\begin{equation}
\Omega_{2}(t)=-{1\over 8}+{1\over 4}t^{2}-{1\over 8}t^{4},
\label{(2.39)}
\end{equation}
\begin{equation}
\omega_{0}(t)={25\over 512}-{531\over 320}t^{2}
+{221 \over 64}t^{4}-{1105 \over 576}t^{6},
\label{(2.40)}
\end{equation}
\begin{equation}
\omega_{1}(t)=-{1\over 16}-{t\over 16}+{5\over 4}t^{2}
+{3\over 8}t^{3}-{49\over 16}t^{4}-{5\over 16}t^{5}
+{15\over 8}t^{6},
\label{(2.41)}
\end{equation}
\begin{equation}
\omega_{2}(t)=-{t\over 16}+{5\over 16}t^{2}+{t^{3}\over 8}
-{5\over 8}t^{4}-{t^{5}\over 16}+{5\over 16}t^{6},
\label{(2.42)}
\end{equation}
\begin{equation}
\omega_{3}(t)=-{1\over 24}+{t^{2}\over 8}-{t^{4}\over 8}
+{t^{6}\over 24}.
\label{(2.43)}
\end{equation}
In the calculation, all factors $1+t$ in the denominators of
$A_{1},A_{2}$ and $A_{3}$ have cancelled against factors in the
numerators, as in the $\theta=0$ case [12]. Moreover, a simple
but non-trivial consistency check shows that, at $\theta=0$,
equations (2.34)--(2.43) yield $A_{1},A_{2}$ and $A_{3}$ in
agreement with Ref. [12].

As a result of all these formulae we find
\begin{equation}
\log \Bigr[(ix)^{-2(\nu-1)}(J_{\nu-1}^{2}-
e^{2\theta}J_{\nu}^{2})(ix)\Bigr]
\sim \sum_{i=1}^{\infty}{\widetilde S}_{i}(\nu,\alpha_{\nu}(x)),
\label{(2.44)}
\end{equation}
where the first few functions ${\widetilde S}_{i}$ read
(cf. Ref. [12])
\begin{equation}
{\widetilde S}_{1} \equiv -\log \pi +2 \alpha_{\nu},
\label{(2.45)}
\end{equation}
\begin{equation}
{\widetilde S}_{2} \equiv -(2\nu-1)\log(\nu+\alpha_{\nu})
+\log f_{\theta}(t),
\label{(2.46)}
\end{equation}
\begin{equation}
{\widetilde S}_{3} \equiv {A_{1}\over \alpha_{\nu}},
\label{(2.47)}
\end{equation}
\begin{equation}
{\widetilde S}_{4} \equiv {A_{2}\over \alpha_{\nu}^{2}},
\label{(2.48)}
\end{equation}
\begin{equation}
{\widetilde S}_{5} \equiv {A_{3}\over \alpha_{\nu}^{3}}.
\label{(2.49)}
\end{equation}
Expanding $\mbox{deg}(n)$ in powers of $n$,
the infinite sum over $n$ in the expression (2.9) can be evaluated
with the help of formulae derived using contour
integration, i.e. \cite{[20]}:
\begin{equation}
\sum_{p=0}^{\infty}p^{2k}\alpha_{p}^{-2k-l}
\sim {\Gamma \left(k+{1\over 2}\right)
\Gamma \left({l\over 2}-{1\over 2}\right)\over
2 \Gamma \left(k+{l\over 2}\right)}x^{1-l}, \;
k=1,2,...,
\label{(2.50)}
\end{equation}
\begin{equation}
\sum_{p=0}^{\infty}p\alpha_{p}^{-1-l} \sim
{x^{1-l}\over \sqrt{\pi}}\sum_{r=0}^{\infty}
{2^{r}\over r!}{\widetilde B}_{r}x^{-r}
{\Gamma \left({r\over 2}+{1\over 2}\right)
\Gamma \left({l\over 2}-{1\over 2}+{r\over 2}\right)
\over 2 \Gamma \left({1\over 2}+{l\over 2}\right)}
\cos {r \pi \over 2}.
\label{(2.51)}
\end{equation}
Here $l$ is a real number larger than $1$ and
${\widetilde B}_{0}=1,{\widetilde B}_{2}={1\over 6},
{\widetilde B}_{4}=-{1\over 30}$ etc. are Bernoulli numbers.
In arbitrary dimension $d$, the expansion of $\mbox{deg}(n)$ in powers
of $n$ is cumbersome and a systematic formula suitable for all $d$ is given
in Eq. (\ref{asymbarnes}).

\section{Conformal anomaly on the 4-ball}

As a first application, we show how the $\zeta(0)$ calculation
of Ref. [12] is extended to our boundary conditions involving
$\theta$, leading in turn to the eigenvalue conditions (1.6)
and (1.7). Only calculations for (1.7) are presented, but the
full $\zeta(0)$ value, which expresses the conformal anomaly
for a massless Dirac spin-${1\over 2}$ field (we do not
study the coupling of spinor fields to gauge fields, which
would lead us instead to the subject of chiral anomalies), receives
a contribution from (1.6) obtained by replacing $\theta$ with
$-\theta$ in the result from (1.7).

In 4 dimensions, Sec. II shows that $\zeta(0)=a_{2}$ is
equal to ${1\over 2}$ times the coefficient of $x^{-6}$ in the
asymptotic expansion of the right-hand side of Eq. (2.9) at
$d=4$. On setting $m \equiv n+2$ the latter reads (here
$\alpha_{m}(x) \equiv \sqrt{m^{2}+x^{2}}$)
\begin{equation}
\sum_{m=0}^{\infty}(m^{2}-m)\left({1\over 2x}{d\over dx}\right)^{3}
\sum_{i=1}^{\infty}{\widetilde S}_{i}(m,\alpha_{m}(x))
\sim \sum_{i=1}^{\infty}W_{\infty}^{i},
\label{(3.1)}
\end{equation}
with $W_{\infty}^{i}$ corresponding to the third derivative
of ${\widetilde S}_{i}$, for all $i$. The terms $W_{\infty}^{i}$
contribute to $a_{2}$ in 4 dimensions only up to $i=5$, and hence
only their analysis is presented hereafter.
\vskip 0.3cm
\centerline{\it Contribution of $W_{\infty}^{1}$ and
$W_{\infty}^{2}$}
\vskip 0.3cm
The term $W_{\infty}^{1}$ is given by
\begin{equation}
W_{\infty}^{1}=\sum_{m=0}^{\infty}(m^{2}-m)
\left({1\over 2x}{d\over dx}\right)^{3}(-\log(\pi)
+2\alpha_{m}),
\label{(3.2)}
\end{equation}
which is unaffected by $\theta$-dependent boundary conditions.
Thus, we know from Ref. [12] that Eqs. (2.50) and (2.51) imply
vanishing contribution to $\zeta(0)$.

The term $W_{\infty}^{2}$ reads (here $t \equiv {m\over \alpha_{m}}$)
\begin{eqnarray}
W_{\infty}^{2}&=&\sum_{m=0}^{\infty}(m^{2}-m)
\left({1\over 2x}{d\over dx}\right)^{3}
\Bigr[-(2m-1)\log(m+\alpha_{m})+\log f_{\theta}(t)\Bigr]
\nonumber \\
&=& W_{\infty}^{2,A}+W_{\infty}^{2,B},
\label{(3.3)}
\end{eqnarray}
where $W_{\infty}^{2,A}$ is the $\theta$-independent part while
$W_{\infty}^{2,B}$ denotes the part involving $\log f_{\theta}(t)$.
>From Ref. [12] we know that $W_{\infty}^{2,A}$ contributes
\begin{equation}
\zeta^{2,A}(0)=-{1\over 120}+{1\over 24}={1\over 30}.
\label{(3.4)}
\end{equation}
The $\log f_{\theta}$ is dealt with by defining
\begin{equation}
\gamma \equiv {1\over 2}(1-e^{2\theta})=-e^{\theta}
\sinh \theta ,
\label{(3.5)}
\end{equation}
\begin{equation}
\beta \equiv {\gamma \over (1-\gamma)}=-\tanh \theta ,
\label{(3.6)}
\end{equation}
and hence writing (see (2.33))
\begin{equation}
\log f_{\theta}=\log (1-\gamma) + \log (1+\beta m \alpha_{m}^{-1}).
\label{(3.7)}
\end{equation}
At this stage we can exploit Eq. (2.28), with
$f \equiv \beta m \alpha_{m}^{-1} < 1$ since $\alpha_{m}$ is
always evaluated at large $x$, and hence we find
\begin{equation}
W_{\infty}^{2,B}={1\over 8}\sum_{k=1}^{\infty}(-1)^{k}
(k+2)(k+4)\beta^{k}\sum_{m=0}^{\infty}(m^{2+k}-m^{1+k})
\alpha_{m}^{-k-6},
\label{(3.8)}
\end{equation}
where the interchange of the orders of summation is made possible
by uniform convergence. Interestingly, this sum contributes
infinitely many $x^{-6}$ terms with equal magnitude and opposite
sign, so that $\zeta(0)$ is unaffected. More precisely, we consider
odd and even values of $k$ and hence define
\begin{equation}
F_{1} \equiv \sum_{m=0}^{\infty}m^{2k+3}\alpha_{m}^{-2k-7}, \;
F_{2} \equiv \sum_{m=0}^{\infty}m^{2k+2}\alpha_{m}^{-2k-7}, \;
k=0,1,2,...,
\label{(3.9)}
\end{equation}
\begin{equation}
F_{3} \equiv \sum_{m=0}^{\infty}m^{2k+2}\alpha_{m}^{-2k-6}, \;
F_{4} \equiv \sum_{m=0}^{\infty}m^{2k+1}\alpha_{m}^{-2k-6}, \;
k=1,2,...
\label{(3.10)}
\end{equation}
By virtue of (2.51), $F_{1}$ contributes to $x^{-6}$ with zero weight
for all $k$ because of the $\cos {3\pi \over 2}$ coefficient.
Moreover, $F_{2}$ is proportional to $x^{-4}$ by virtue of (2.50).
The sum $F_{3}$ is instead proportional to $x^{-3}$ (again by (2.50)),
while $F_{4}$ is such that its contribution
$\delta_{F_{4}}(x;k)$ to $\zeta(0)$ reads
\begin{equation}
\delta_{F_{4}}(x;1)=-{1\over 12}x^{-6}\left({\Gamma(3)\over
\Gamma(3)}-{\Gamma(4)\over \Gamma(4)}\right)=0,
\label{(3.11)}
\end{equation}
\begin{equation}
\delta_{F_{4}}(x;2)-\delta_{F_{4}}(x;1)
=-{1\over 12}x^{-6}\left(-{\Gamma(4)\over \Gamma(4)}
+{\Gamma(5)\over \Gamma(5)}\right)=0,
\label{(3.12)}
\end{equation}
and infinitely many other relations along the same lines.
\vskip 0.3cm
\centerline {\it Effect of $W_{\infty}^{3},W_{\infty}^{4}$ and
$W_{\infty}^{5}$}
\vskip 0.3cm
The term $W_{\infty}^{3}$ is equal to
\begin{eqnarray}
W_{\infty}^{3}&=& \sum_{m=0}^{\infty}(m^{2}-m)
\left({1\over 2x}{d\over dx}\right)^{3}\biggr[{1\over 4}
\alpha_{m}^{-1}-{5\over 12}m^{2}\alpha_{m}^{-3} \nonumber \\
&+& {1\over 2}(1-\gamma)^{-1}\left({m^{2}\over \alpha_{m}^{3}}
-{1\over \alpha_{m}}\right)(1+\beta m \alpha_{m}^{-1})^{-1}
\biggr] \nonumber \\
&=&W_{\infty}^{3,A}+W_{\infty}^{3,B}+W_{\infty}^{3,C}
+W_{\infty}^{3,D},
\label{(3.13)}
\end{eqnarray}
where $W_{\infty}^{3,A}$ and $W_{\infty}^{3,B}$ are the first
two, $\theta$-independent sums, while $W_{\infty}^{3,C}$ and
$W_{\infty}^{3,D}$ are the sums depending on $\theta$ through
$\gamma$ and $\beta$. By virtue of the large-$x$ nature of the
whole analysis, we can expand $(1+\beta m \alpha_{m}^{-1})^{-1}$
according to
\begin{equation}
(1+\beta m \alpha_{m}^{-1})^{-1}=\sum_{k=0}^{\infty}
(-1)^{k}\beta^{k}m^{k}\alpha_{m}^{-k}.
\label{(3.14)}
\end{equation}
Upon exploiting the identity
\begin{equation}
\left({1\over 2x}{d\over dx}\right)^{3}\alpha_{m}^{-l}
=-{1\over 8}l(l+2)(l+4)\alpha_{m}^{-l-6},
\label{(3.15)}
\end{equation}
we find that $W_{\infty}^{3,A}$ and $W_{\infty}^{3,B}$ do not
contribute to $\zeta(0)$ by virtue of (2.50) and (2.51). The
same holds for $W_{\infty}^{3,C}$ and $W_{\infty}^{3,D}$, but
the proof requires more intermediate steps, as follows. The
term $W_{\infty}^{3,C}$ is given by
\begin{equation}
W_{\infty}^{3,C}=-{1\over 16}(1-\gamma)^{-1}\sum_{k=0}^{\infty}
(-1)^{k}\beta^{k}(k+3)(k+5)(k+7)\sum_{m=0}^{\infty}
(m^{4}-m^{3})m^{k}\alpha_{m}^{-k-9}.
\label{(3.16)}
\end{equation}
Looking at even and odd values of $k$, this suggests defining
\begin{equation}
G_{1} \equiv \sum_{m=0}^{\infty}m^{2k+4}\alpha_{m}^{-2k-9}, \;
G_{2} \equiv \sum_{m=0}^{\infty}m^{2k+3}\alpha_{m}^{-2k-9}, \;
k=0,1,2 ...,
\label{(3.17)}
\end{equation}
\begin{equation}
G_{3} \equiv \sum_{m=0}^{\infty} m^{2k+5}\alpha_{m}^{-2k-10}, \;
G_{4} \equiv \sum_{m=0}^{\infty}m^{2k+4}\alpha_{m}^{-2k-10}, \;
k=0,1,2 ... \; .
\label{(3.18)}
\end{equation}
Now $G_{1}$ and $G_{4}$ are proportional to $x^{-4}$ and $x^{-5}$
respectively by virtue of (2.50), and hence do not contribute to
$\zeta(0)$. $G_{2}$ contains $x^{-6}$ weighted by a coefficient
proportional to $\cos {\pi \over 2}$, for all $k$, and hence does
not contribute to $\zeta(0)$. Last, $G_{3}$ is such that its
contribution $\delta_{G_{3}}(x;k)$ to $\zeta(0)$ reads
\begin{equation}
\delta_{G_{3}}(x;0)=-{1\over 12}{x^{-6}} \left(
{\Gamma(5/2)\over \Gamma(5/2)}
-2{\Gamma(4)\over \Gamma(4)}
+{\Gamma(5)\over \Gamma(5)}\right)=0,
\label{(3.19)}
\end{equation}
jointly with infinitely many other relations along the same lines.

The term $W_{\infty}^{3,D}$ is given by
\begin{equation}
W_{\infty}^{3,D}={1\over 16}(1-\gamma)^{-1}\sum_{k=0}^{\infty}
(-1)^{k}\beta^{k}(k+1)(k+3)(k+5)\sum_{m=0}^{\infty}
(m^{2+k}-m^{1+k})\alpha_{m}^{-k-7}.
\label{(3.20)}
\end{equation}
Here, too, we split the sum over $k$ into sums over all even
and odd values of $k$. We find therefore, exploiting (2.50)
and (2.51), either contributions proportional to $x^{-4}$ and
$x^{-5}$, or $x^{-6}$ terms weighted by $\cos {\pi \over 2}$,
or the contributions resulting from
\begin{equation}
H_{3} \equiv \sum_{m=0}^{\infty}m^{2k+3}\alpha_{m}^{-2k-8},
\; k=0,1,2,...,
\label{(3.21)}
\end{equation}
which occur with opposite signs for all $k$.

The general formula for $W_{\infty}^{4}$ reads
\begin{eqnarray}
W_{\infty}^{4}&=& \sum_{m=0}^{\infty}(m^{2}-m)
\left({1\over 2x}{d\over dx}\right)^{3}{A_{2}\over \alpha_{m}^{2}}
\nonumber \\
&=& \sum_{m=0}^{\infty}(m^{2}-m)\left({1\over 2x}
{d\over dx}\right)^{3}\left \{ \alpha_{m}^{-2}
\biggr[\sum_{r=0}^{2}j_{0,r}m^{2r}\alpha_{m}^{-2r}
\right . \nonumber \\
&+& \left . f_{\theta}^{-1}\sum_{r=1}^{4}j_{1,r}m^{r}\alpha_{m}^{-r}
+ f_{\theta}^{-2}\sum_{r=0}^{2}j_{2,r}m^{2r}
\alpha_{m}^{-2r}\Bigr] \right \} \nonumber \\
&=& W_{\infty}^{4,A}+W_{\infty}^{4,B}+W_{\infty}^{4,C},
\label{(3.22)}
\end{eqnarray}
where $j_{0,r},j_{1,r}$ and $j_{2,r}$ are the coefficients in
the polynomials $\Omega_{0},\Omega_{1}$ and $\Omega_{2}$
respectively (see (2.37)--(2.39)), and negative powers of
$f_{\theta}$ are expanded by exploiting
\begin{equation}
(1+f)^{-s}=\sum_{k=0}^{\infty}(-1)^{k}
{\Gamma(k+s)\over k! \Gamma(s)}f^{k} \;
{\rm as} \; f \rightarrow 0.
\label{(3.23)}
\end{equation}
Since
\begin{equation}
W_{\infty}^{4,A}=-{1\over 8}\sum_{r=0}^{2}j_{0,r}
(2r+2)(2r+4)(2r+6)\sum_{m=0}^{\infty}
(m^{2r+2}-m^{2r+1})\alpha_{m}^{-2r-8},
\label{(3.24)}
\end{equation}
the basic formulae (2.50) and (2.51) imply a contribution
to $\zeta(0)$ equal to
\begin{equation}
{1\over 2}\sum_{r=0}^{2}j_{0,r}=0.
\label{(3.25)}
\end{equation}
Moreover, since
\begin{eqnarray}
W_{\infty}^{4,B}&=& -{1\over 8}(1-\gamma)^{-1}
\sum_{r=1}^{4}j_{1,r}\sum_{k=0}^{\infty}(-1)^{k}\beta^{k}
(k+r+2)(k+r+4)(k+r+6) \nonumber \\
& \times & \sum_{m=0}^{\infty}(m^{2}-m)m^{k+r}
\alpha_{m}^{-k-r-8} ,
\label{(3.26)}
\end{eqnarray}
\begin{eqnarray}
W_{\infty}^{4,C}&=& -{1\over 8}(1-\gamma)^{-2}
\sum_{r=0}^{4}j_{2,r}\sum_{k=0}^{\infty}(-1)^{k}
(k+1)\beta^{k}(k+2r+2)(k+2r+4)(k+2r+6) \nonumber \\
& \times & \sum_{m=0}^{\infty}(m^{2(r+1)+k}
-m^{2r+k+1})\alpha_{m}^{-k-2r-8},
\label{(3.27)}
\end{eqnarray}
repeated application of (2.50) and (2.51) yields contributions
to $\zeta(0)$ equal to
\begin{equation}
{1\over 2}(1-\gamma)^{-1}(1+\beta)^{-1}\sum_{r=1}^{4}
j_{1,r}=0,
\label{(3.28)}
\end{equation}
and
\begin{equation}
{1\over 2}(1-\gamma)^{-2}\left(\sum_{k=0}^{\infty}
(-1)^{k}(k+1)\beta^{k}\right)\sum_{r=0}^{2}j_{2,r}=0,
\label{(3.29)}
\end{equation}
respectively. The results (3.25), (3.28) and (3.29) are
all vanishing because of the peculiar properties of the
$j_{0,r},j_{1,r}$ and $j_{2,r}$ coefficients.

Last, the general formula for $W_{\infty}^{5}$ reads
\begin{eqnarray}
W_{\infty}^{5}&=& \sum_{m=0}^{\infty}(m^{2}-m)
\left({1\over 2x}{d\over dx}\right)^{3}
{A_{3}\over \alpha_{m}^{3}} \nonumber \\
&=& \sum_{m=0}^{\infty}(m^{2}-m)
\left({1\over 2x}{d\over dx}\right)^{3}
\left \{ \alpha_{m}^{-3} \biggr[\sum_{r=0}^{3}
\sigma_{0,2r}m^{2r}\alpha_{m}^{-2r} \right . \nonumber \\
&+& f_{\theta}^{-1}\sum_{r=0}^{6}\sigma_{1,r}
m^{r}\alpha_{m}^{-r}+f_{\theta}^{-2}
\sum_{r=1}^{6}\sigma_{2,r}m^{r}\alpha_{m}^{-r}
\nonumber \\
&+& \left . f_{\theta}^{-3}\sum_{r=0}^{3}\sigma_{3,2r}
m^{2r}\alpha_{m}^{-2r} \Bigr] \right \} \nonumber \\
&=& W_{\infty}^{5,A}+W_{\infty}^{5,B}+W_{\infty}^{5,C}
+W_{\infty}^{5,D},
\label{(3.30)}
\end{eqnarray}
where $\sigma_{0,2r},\sigma_{1,r},\sigma_{2,r}$ and
$\sigma_{3,2r}$ are the coefficients in the polynomials
$\omega_{0},\omega_{1},\omega_{2}$ and $\omega_{3}$
respectively (see (2.40)--(2.43)), and also $f_{\theta}^{-3}$
is expanded by exploiting (3.23). Now the term
$W_{\infty}^{5,A}$, which is the $\theta$-independent part
of (3.30), yields a non-vanishing contribution to $\zeta(0)$
equal to
\begin{equation}
{1\over 2}\left(-\sum_{r=0}^{3}\sigma_{0,2r}\right)
=-{1\over 360},
\label{(3.31)}
\end{equation}
while the terms $W_{\infty}^{5,B},W_{\infty}^{5,C}$ and
$W_{\infty}^{5,D}$, resulting from $f_{\theta}^{-1},
f_{\theta}^{-2}$ and $f_{\theta}^{-3}$ respectively, give
vanishing contribution obtained as follows:
\begin{equation}
{1\over 2}(1-\gamma)^{-1}(1+\beta)^{-1}
\left(-\sum_{r=0}^{6}\sigma_{1,r}\right)=0 \;
{\rm from} \; W_{\infty}^{5,B},
\label{(3.32)}
\end{equation}
\begin{equation}
{1\over 2}(1-\gamma)^{-2}\left(\sum_{k=0}^{\infty}
(-1)^{k}(k+1)\beta^{k}\right)
\left(-\sum_{r=1}^{6}\sigma_{2,r}\right)=0 \;
{\rm from} \; W_{\infty}^{5,C},
\label{(3.33)}
\end{equation}
\begin{equation}
{1\over 2}(1-\gamma)^{-3}\left(\sum_{k=0}^{\infty}
(-1)^{k}{(k+1)(k+2)\over 2}\beta^{k}\right)
\left(-\sum_{r=0}^{3}\sigma_{3,2r}\right)=0 \;
{\rm from} \; W_{\infty}^{5,D},
\label{(3.34)}
\end{equation}
by repeated application of (2.50) and (2.51).

By virtue of (3.4), (3.25), (3.28), (3.29),
(3.31)--(3.34) we find
\begin{equation}
\zeta(0)=2 \left({1\over 30}-{1\over 360}\right)
={11\over 180},
\label{(3.35)}
\end{equation}
for a massless Dirac field on the 4-ball, bearing in
mind that also the eigenvalue condition (1.6) should be
considered. Interestingly, such a $\zeta(0)$ value in 4
dimensions is independent of $\theta$, and agrees with the
result in Ref. [12], where a massless spin-${1\over 2}$ field
with half as many components as a Dirac field was instead
considered.

The proof of vanishing contributions to $\zeta(0)$ from the infinite
sums $F_{4},G_{3}$ and $H_{3}$ can be made more systematic and
elegant by remarking that a recursive scheme exists for which
\begin{equation}
H_{3}=\left(1+{x\over 2(k+3)}{d\over dx}\right)F_{4},
\label{(3.36)}
\end{equation}
\begin{equation}
G_{3}=\left(1+{x\over 2(k+4)}{d\over dx}\right)H_{3},
\label{(3.37)}
\end{equation}
so that one only needs to look at $F_{4}$ (see (3.10)),
which can be evaluated
exactly as a function of $x$ for all $k$ by exploiting the
Euler--Maclaurin formula [3,12].

\section{Heat-kernel coefficients in general dimension $d$}

Our aim in this section is to apply the formalism in such a way
that in principle all heat-kernel coefficients in any
dimension $d$ can be obtained.
Therefore we will need the large-$x$ expansion of
(see (2.21) and (3.5))
\beq
\left(\frac 1 {2x} \frac d {dx} \right)^{1+d/2} \left[
 2\anu -( 2\nu -1 ) \log (\nu +\anu )
 +\log (1+\gamma (t-1)) +
\sum_{p=1}^\infty \frac{A_p}{(\anu )^p} \right].
\label{(4.1)}
\eeq
The first complication compared to $d=4$ is that now, dealing with
arbitrary dimension $d$,
we need an arbitrary number of derivatives. This is
easily generalized in some cases, i.e.
\beq
\dj \anu &=& (-1)^{j+1} \frac{(2j-3)!!}{2^j} \anu ^{1-2j},\nn
\eeq
in others at least in the form of a large-$\anu$ expansion,
\beq
\dj \log (\nu +\anu ) & \sim &
\frac 1 2 (-1)^{j+1} \Gamma (j) \anu^{-2j} \nn\\
  & &+\frac {(-1)^j}{2^j} \sum_{k=1}^\infty (-1)^{k+1} \frac{(k+2j-2)!!}
{k!!} \nu^k \anu ^{-k-2j} ,\nn\\
\dj \log (1+\gamma (t-1)) & \sim &
\frac {(-1)^j}{2^j} \sum_{k=1}^\infty
 (-1)^{k+1}
\frac{(k+2j-2)!!}
{k!!} \nu^k \beta^k \anu ^{-k-2j} .\nn
\eeq
As we have seen in Secs. II and III,
to deal with the $A_p$ contributions, we need terms of the type
\beq
\lefteqn{
\dj \frac{t^i} {(1+\gamma (t-1)^l) \anu ^p} =\frac{(-1)^j}{2^j}\frac 1
{(1-\gamma)^l} }\nn\\
 & & \quad \quad
   \times \sum_{u=0}^\infty (-1)^u \frac{\Gamma (l+u)}{u! \Gamma (l)}
  \beta^u \nu^{u+i} \frac{ (u+i+p+2j -2)!!}{(u+i+p-2)!!} \anu ^{-u-i-p-2j}.
\nn
\eeq
The relevant case is $j=1+d/2$ and the contribution of each term
to the zeta function is found by summing over $n$, taking the degeneracy
into account.

Let us now show how the general procedure works in the case
of the $ \anu$-term. The
contribution to $\zeta (1+d/2,x^2)$ is
\beq
B &=& (-1)^{d/2} \frac {d_s}{2 \Gamma (1+d/2)} \sum_{n=0}^\infty
\bn \left( \frac 1 {2x} \frac d {dx}\right)^{1+d/2}  2 \anu \nn\\
 &=& d_s \frac{(d-1)!!}{2^{1+d/2}
   \Gamma \left( 1+\frac d 2 \right)} \sum_{n=0}^\infty
\bn  \anu ^{-1-d}\nn\\
 &=& d_s\frac{(d-1)!!}{2^{1+d/2}\Gamma \left( 1+\frac d 2 \right)}
   \sum_{n=0}^\infty
\bn (\nu^2 +x^2 )^{-\frac {1+d} 2} ,
\label{(4.2)}
\eeq
of which we need the large-$x$ expansion. A simple expansion in inverse
powers of $x$ is not allowed; instead we employ a
Mellin integral representation. As is clear from the above equation, the
Barnes zeta function
\cite{barn03-19-426,barn03-19-374,dowk94-35-4989,dowk94-162-633}
\beq
\zeta_{{\cal B}} (s,a) \equiv \sum_{n=0}^\infty
\bn (n+a)^{-s}
 = \sum_{\vec m =0} ^\infty (a+m_1 + ... + m_{d-1} ) ^{-s} , \nn
\eeq
will play a crucial role. We need to separate the $\nu$ and $x$ dependence
in (4.2), more generally in expressions of the form
\beq
C(j,s) & :=  &\sum_{n=0}^\infty \bn \nu^j \anu ^{-j-s} \nn\\
   &=& \frac 1 {\Gamma \left( \frac{s+j} 2 \right)} \sum_{n=0}^\infty
  \bn \nu^j \int_0^\infty dt \;
t^{\frac{s+j} 2 -1} e^{-(\nu^2 +x^2) t} .\nn
\eeq
The $\nu$ and $x$ dependence is separated by employing for $\Re c >0$,
\beq
e^{-\nu ^2 t} = \frac 1 {2\pi i} \int\limits_{c-i\infty}^{c+i\infty}
d\alpha \Gamma (\alpha ) \nu ^{-2\alpha } t^{-\alpha} .\nn
\eeq
For $\Re s$ large enough we continue
\beq
C( j,s) &=& \frac 1 {\Gamma \left( \frac{s+j} 2 \right) } \sum_{n=0}^\infty
\bn \nu^j \frac 1 {2\pi i} \int\limits_{c-i\infty}^{c+i\infty}
d\alpha \Gamma (\alpha ) \nu^{-2\alpha}
   \int\limits_0^\infty dt \; t^{\frac{s+j} 2 -\alpha -1}
  e^{-x^2 t} \nn\\
   &=&  \frac 1 {\Gamma \left( \frac{s+j} 2 \right) } \sum_{n=0}^\infty
\bn \nu^j \frac 1 {2\pi i} \int\limits_{c-i\infty}^{c+i\infty}
d\alpha \Gamma (\alpha ) \nu^{-2\alpha} \Gamma \left( \frac{s+j} 2 -\alpha
\right) x^{2\alpha -s-j} .\nn
\eeq
The sum and integral may be interchanged upon choosing
$\Re c > (j+d-1)/2$ and we find
\beq
C(j,s) =  \frac 1 {\Gamma \left( \frac{s+j} 2 \right) }
\frac 1 {2\pi i} \int\limits_{c-i\infty}^{c+i\infty}
d\alpha \Gamma (\alpha )  \Gamma \left( \frac{s+j} 2 -\alpha
\right) x^{2\alpha -s-j} \zeta_{{\cal B}} \left(2\alpha -j , \frac d 2
\right) .\nn
\eeq
On shifting the contour to the left we pick up the large-$x$
expansion of $C(j,s)$. In order to find $a_{n/2}$, we are interested
in the term that behaves as $x^{-n-2}$ and need to evaluate the residue at
$a=(s+j-n-2)/2$. In all cases we encounter, the only relevant pole
will come from $\zeta_{{\cal B}}$ and for these cases
\beq
C(j,s) \sim \frac{\Gamma \left( \frac{s+j-n} 2 -1 \right)  }
 {2 \Gamma \left( \frac { s+j} 2 \right) } \Gamma \left( 1+\frac n 2
\right) \mbox{ Res } \zeta _{{\cal B}} \left( s-n-2 , \frac d 2 \right)
  x^{-n-2} + \mbox{ irrelevant } .\label{asymbarnes}
\eeq
>From here, e.g., one obtains
\beq
B= \frac 1 {4\sqrt{\pi}} d_s \Gamma \left( \frac {d-1-n} 2 \right)
 \mbox{ Res } \zeta _{{\cal B}} \left(d-1-n, \frac d 2 \right)
  \frac {\Gamma \left( 1+\frac n 2 \right) }{\Gamma
\left(1+\frac d 2 \right) } .
\nn
\eeq
The procedure just outlined can be applied to all terms in (4.1).
The following list summarizes for each term on the left the contribution
to the heat-kernel coefficient $a_{n/2}$ on the right ($\theta$ and $-\theta$
contributions are summed):
\beq
(2\anu -2\nu \log (\nu + \anu )) &\to &\frac{ d_s} {2\sqrt{\pi} (d-n)}
  \Gamma \left( \frac{ d-n-1} 2 \right) \mbox{ Res } \zeta_{{\cal B}}
   \left( d-1-n , \frac d 2 \right) , \nn\\
\log (\nu + \anu) &\to & \frac{d_s} {2\sqrt{\pi} (d-n)} \Gamma
\left( \frac{d-n+1} 2 \right)  \mbox{ Res } \zeta_{{\cal B}}
   \left( d-n , \frac d 2 \right) , \nn\\
\log (1+\gamma (t-1)) &\to & \frac {d_s} 4 \Gamma \left( \frac{d-n} 2 \right)
  \left( \cosh ^{d-n} \theta -1 \right)
 \mbox{ Res } \zeta_{{\cal B}}
   \left( d-n , \frac d 2 \right) , \nn\\
\frac{t^i} {\anu^p (1+\gamma (t-1))^l} &\to & -\frac{d_s} {4e^{l\theta}
\cosh ^l \theta } \mbox{ Res } \zeta_{{\cal B}}
   \left( d+p-n , \frac d 2 \right) \times\nn\\
& &\hspace{-2.5cm}
\left\{ \frac{\Gamma \left( \frac{ d+i+p-n} 2 \right) }
   {\Gamma \left( \frac{ i+p} 2 \right) }
    {}_3 F_2 \left( \frac{ l+1} 2 , \frac l 2 , \frac{ d+i+p-n} 2 ;
\frac 1 2 , \frac{p+i} 2 ; \tanh ^2 \theta \right) \right. \nn\\
& &\left. \hspace{-2.5cm}
+ l\tanh \theta \frac{\Gamma \left( \frac{d+i+p+1-n} 2 \right) }
   {\Gamma \left( \frac{ i+p+1} 2 \right)} {}_3 F_2 \left(
   \frac{ l+1} 2 , 1+\frac l 2 , \frac{ d+i+p+1-n} 2 ; \frac 3 2 , \frac{
  p+i+1 } 2 ; \tanh ^2 \theta \right) \right\}\nn\\
  & & + \quad \theta \to -\theta  .\nn
\eeq
For the calculation of heat-kernel coefficients, note that only a
finite number of terms contributes. The poles of the Barnes zeta function
are located at $s=1,...,d-1$, and depending on the values of $n$ and $p$
only a finite number of terms needs to be evaluated. In general, to evaluate
$a_{n/2}$, we need to include all terms up to $p=n-1$ [12].

The above results resemble very much the structure of the results found
for
different boundary conditions given in
\cite{dowk96-13-585,bord96-182-371,dowk99-7-641,dowk99-16-1917,dowk9507096}.
In particular,
a reduction of the analysis from the ball to the sphere (in form of
the Barnes zeta functions) has been achieved. Indeed, instead of using
the presented algorithm we could equally well have used the contour
integral method developed in
\cite{bord96-37-895,bord96-182-371,dowk99-7-641}. The starting point for
the zeta function associated with the eigenvalues from (1.7) in this
approach reads
\beq
\zeta_\theta (s) = \sum_{n=0}^\infty \mbox{deg}(n) \int\limits_\gamma
\frac{dk}{2\pi i} k^{-2s} \frac \partial {\partial k}
\ln \left( J_{n+d/2-1} ^2 (k) - e^{2\theta} J_{n+d/2} ^2 (k) \right) ,
\nn
\eeq
the contour $\gamma$ enclosing all eigenvalues of (1.7).
One then uses the uniform asymptotic expansion in order to
extract the pieces that can contribute to the heat kernel coefficients.
Performing the $k$-integrals, results analogous to the above are found
and final answers, of course, agree.

Given the explicit
results in the above list where all ingredients are known,
the algorithm can be cast in a form suitable for application
of Mathematica. As far as this process is concerned, some
remarks are in order. We have presented the results in terms of
hypergeometric functions, and as far as we can see keeping $d,n$
{\it arbitrary} this is the best one can do. However, as soon as one
considers particular values of $d$ and $n$, the hypergeometric function
${}_3F_2$ ``collapses'' to ${}_2F_1$,
which, at the particular values needed,
is simply given as an algebraic
combination of hyperbolic functions. For example one has
\beq
{}_2F_1 (1,1,1/2,\tanh ^2 \theta ) &=& \frac 1 {1-\tanh ^2 \theta }
  + \frac{\tanh \theta \arcsin \tanh \theta }{(1-\tanh ^2 \theta )^{3/2}}
\nn\\
&=& \cosh ^2 \theta (1+\arcsin \tanh \theta  \sinh \theta ). \nn
\eeq
Mathematica will not always replace automatically the hypergeometric
functions by this kind of hyperbolic combinations. Since this is essential
for further simplifications of final answers, the implementation of
some of the Gauss relations is necessary. We have used
\beq
{}_2F_1 (\alpha +1 , \beta +1, \gamma +1 , z) &=& \frac 1 {\alpha (1-z)}
  \left\{ \gamma \,\,{}_2F_1 (\alpha , \beta , \gamma ,z ) - (\gamma -\alpha )
 \,\,  {}_2 F _1 (\alpha , \beta +1 , \gamma +1 , z) \right\} , \nn\\
\gamma \,\, {}_2 F_1 (\alpha , \beta , \gamma, z) &=& (\gamma -\alpha )\,\,
  {}_2 F _1 (\alpha , \beta , \gamma +1 , z) + \alpha\,\,
{}_2 F_1 (\alpha +1 ,
\beta , \gamma +1 , z) .\nn
\eeq
These relations guarantee that ultimately all hypergeometric functions
are given in very explicit terms and that huge simplifications can be
performed explicitly. In Sec. V we have summarized our findings
in $d=2,4,6$ dimensions giving final results up to the coefficient $a_{d/2}$.

\section{List of heat-kernel coefficients}

We list hereafter the general results we have obtained. The lower
coefficients have quite a simple form in all dimensions and the leading
three coefficients are as follows:
\beq
a_0 &=& \frac{d_s}{2^d \Gamma \left( 1+\frac d 2 \right)},
\label{(5.1)}
\eeq
\beq
a_{1/2}= \frac{\sqrt \pi d_s}{2^d \Gamma \left( \frac d 2 \right) }
   \left( (\cosh \theta)^{d-1} -1 \right) ,
\label{(5.2)}
\eeq
\begin{eqnarray}
a_1 &=& \frac{(2d-5) d_s}{3 \,\,\,2^d \Gamma \left( \frac d 2 \right) }
+ \frac{d_s}{2^d \Gamma \left( \frac d 2 \right)} \left\{
    {}_2 F_1 \left( 1 , \frac{d-1} 2 ; \frac 1 2 ; (\tanh \theta )^2
        \right) \right. \nonumber \\
&\; & \left.
      - (d-1) \,\, {}_2F_1 \left( 1 , \frac{d+1} 2 ; \frac 3 2 ;
       (\tanh \theta )^2
    \right) \right\} .
\label{(5.3)}
\end{eqnarray}
Moreover, to show the applicability of our algorithms to arbitrary
dimensions and in principle to any coefficient we give the following
collection of results.\\
{\bf d=2:}
\beq
a_0 = \frac{d_s} 4 ,
\label{(5.4)}
\eeq
\beq
a_{1/2} = \frac{\sqrt \pi d_s} 4 \left( \cosh \theta -1\right),
\label{(5.5)}
\eeq
\beq
a_1 = -\frac{d_s} {12} .
\label{(5.6)}
\eeq
{\bf d=4:}
\beq
a_0 = \frac{d_s} {32} ,
\label{(5.7)}
\eeq
\beq
a_{1/2}= \frac{\sqrt \pi d_s}{16} \left( (\cosh\theta)^3 -1\right),
\label{(5.8)}
\eeq
\beq
a_1 = -\frac{d_s}{16} \cosh 2\theta ,
\label{(5.9)}
\eeq
\beq
a_{3/2} =\frac{\sqrt\pi d_s}{4096} \left(\mbox{sech}
\left(\frac \theta 2 \right)
\right)^4 \left(15+20 \cosh \theta -11 \cosh 2\theta \right),
\label{(5.10)}
\eeq
\beq
a_2 = \frac{11d_s} {720}.
\label{(5.11)}
\eeq
{\bf d=6:}
\beq
a_0 = \frac{d_s}{384},
\label{(5.12)}
\eeq
\beq
a_{1/2} = \frac{\sqrt\pi d_s}{128} \left( (\cosh \theta )^5 -1\right),
\label{(5.13)}
\eeq
\beq
a_{1} = -\frac{d_s}{384} \left( -2+6\cosh 2\theta + \cosh 4\theta \right),
\label{(5.14)}
\eeq
\beq
a_{3/2} = \frac{\sqrt \pi d_s} {98304} \left(\mbox{sech}
\left(\frac \theta 2
\right)\right)^4 \left( 153 + 212 \cosh \theta + 35 \cosh 2\theta -32
    \cosh 4\theta - 8 \cosh 5\theta \right) ,
\label{(5.15)}
\eeq
\beq
a_2 = \frac {d_s}{96} \cosh 2\theta ,
\label{(5.16)}
\eeq
\begin{eqnarray}
a_{5/2}& =& -\frac{\sqrt \pi d_s} {805306368}
\left(\mbox{sech} \left(\frac \theta 2 \right) \right)^{10}
\left(
311902 + 495474 \cosh \theta + 172792 \cosh 2\theta \right.\nn\\
& &\phantom{-\frac{\sqrt \pi d_s} {805306368}}\left.
  +14845 \cosh 3\theta - 21590 \cosh 4\theta -2159 \cosh 5\theta \right),
\label{(5.17)}
\end{eqnarray}
\beq
a_3 = -\frac{191 d_s}{60480} .
\label{(5.18)}
\eeq

Of course, the result (5.11) for $a_{2}$ in dimension four agrees with
Eq. (3.35), upon bearing in mind that $d_{s}$ is then equal to $4$.
For $\theta =0$ the results agree with the results found previously
in \cite{[18],kirs96-13-633,dowk99-7-641}.

\section{Concluding remarks}

Motivated by quantum cosmology and the problems of quark confinement,
we have studied heat-kernel asymptotics for the squared Dirac
operator on the Euclidean ball, with local boundary conditions (1.1)
leading to the eigenvalue conditions (1.6) and (1.7). We have first
proved that on the 4-ball the $\zeta(0)$ value is $\theta$-independent.
Furthermore,
arbitrary values of $d$ have been considered, and several explicit
formulae for heat-kernel coefficients in dimension $d=2,4,6$ have
been obtained in Secs. IV and V. Interestingly, $a_{d/2}$ is always
$\theta$-independent,
while several other heat-kernel coefficients depend
on $\theta$ through hyperbolic functions and their integer powers.

As far as we can see, the key task is now the analysis of heat-kernel
asymptotics with local boundary conditions (1.1) on general
Riemannian manifolds $(M,g)$ with boundary $\partial M$. One has
then to consider the smooth function $f \in C^{\infty}(M)$ mentioned
after Eq. (1.13), which is replaced by
\begin{equation}
a_{n/2}(f,P,{\cal B})=c_{n/2}(f,P)+b_{n/2}(f,P,{\cal B}).
\label{(6.1)}
\end{equation}
The interior part $c_{n/2}$ vanishes for all odd values of $n$,
whereas the boundary part only vanishes if $n=0$. The interior
part is obtained by integrating over $M$ a linear combination
of local invariants of the appropriate dimension, where the
coefficients of the linear combination are {\it universal constants},
independent of $d$. Moreover, the boundary part $b_{n/2}$ is
obtained upon integration over $\partial M$ of another linear
combination of local invariants. In that case, however, the structure
group is $O(d-1)$, and the coefficients of linear combination will depend
on $d$ and $\theta$ \cite{[17]} and so they will be
universal functions, as it happens if the boundary operator involves
tangential derivatives
\cite{dowk99-16-1917,avit91-8-1445,avra98-15-281,eliz99-16-813}. This
is indeed the case for the boundary condition
(1.1). To see this define
\beq
\chi \equiv i e^{\theta \gamma^5} \gamma^5 \gamma_m\nn
\eeq
and introduce the ``projections''
\beq
\Pi _\pm = \frac 1 2 \left( 1 \pm \chi \right). \nn
\eeq
In the bulk of our article we considered the operator $P$ with domain
\beq
\mbox{domain}(P)= \left\{ \psi \in C^\infty (V) : \Pi _- \psi \left|_{
\partial M} \oplus \Pi _- D \psi \right| _{\partial M} =0 \right\}.\nn
\eeq
We calculate that
\beq
\Pi _- D \psi \left|_{
\partial M} = \left( \Pi _+ ^* \nabla _m + \gamma _m \gamma_a \Pi _- ^*
\nabla _a \right) \Pi _+ \psi  \right| _{\partial M},\nn
\eeq
$a$ being a tangential index, which for Hermitean $\Pi _\pm$
($\theta =0$) would reduce
to standard mixed boundary conditions. However, as is easily seen,
this is not the case for $\theta \neq 0$
and tangential derivatives occur in the
boundary conditions such that the boundary conditions considered could be
termed of mixed oblique type.
It is thus expected, that
the
general form of $a_{n/2}$ contains all possible local invariants
built from $f$, Riemann curvature $R_{\; bcd}^{a}$ of $M$, bundle
curvature $\Omega_{ab}$ (in case a gauge theory, with vector bundle
over $M$, is studied), extrinsic curvature $K_{ij}$ of $\partial M$,
endomorphism $E$ (i.e. potential term) coming from the differential operator
$P$, combinations of $\gamma$-matrices coming from the boundary operator, and
the covariant derivatives of all these geometric objects, eventually
integrating their linear combinations
over $M$ and $\partial M$ \cite{[5],[8],[9],[22]}. All these local invariants
are multiplied by universal functions which might depend on $d$ and $\theta$.
As a next step, the presented special case calculation together with
various other ingredients such as conformal transformations
\cite{bran90-15-245}, index theory \cite{bran99-563-603}, redefinition
of the covariant derivative \cite{spectralII} will serve to find
results valid for arbitrary Riemannian manifolds and bundle curvatures.

\acknowledgments

The work of G. Esposito has been partially supported by PRIN 2000
``Sintesi.'' K. Kirsten thanks Peter Gilkey
for interesting discussions. The work of K. Kirsten
has been supported by the INFN and by the Max-Planck-Institute
for Mathematics in the Sciences, Leipzig. G. Esposito remains
much indebted to Ian Moss for having introduced him to the work
in Ref. \cite{[20]} a long time ago.

\appendix
\section*{}

The function $\cal F$ on the left-hand side of Eq. (1.7) is the product
of the entire functions (i.e. functions analytic in the whole
complex plane)
$$
{\cal F}_{1} \equiv J_{n+d/2-1}-e^{\theta}J_{n+d/2} \; \; {\rm and} \; \;
{\cal F}_{2} \equiv J_{n+d/2-1}+e^{\theta}J_{n+d/2},
$$
which can be written in the form
\begin{equation}
{\cal F}_{1}(k)=\gamma_{1}k^{n+d/2-1}e^{g_{1}(k)}
\prod_{i=1}^{\infty}\left(1-{k\over \mu_{i}}\right)
e^{k \over \mu_{i}},
\label{(A1)}
\end{equation}
\begin{equation}
{\cal F}_{2}(k)=\gamma_{2}k^{n+d/2-1}e^{g_{2}(k)}
\prod_{i=1}^{\infty}\left(1-{k\over \nu_{i}}\right)
e^{k \over \nu_{i}}.
\label{(A2)}
\end{equation}
In Eqs. (A1) and (A2), $\gamma_{1}$ and $\gamma_{2}$ are constants,
$g_{1}$ and $g_{2}$ are entire functions, the $\mu_{i}$ are the
zeros of ${\cal F}_{1}$ and the $\nu_{i}$
are the zeros of ${\cal F}_{2}$. The
general theory described in Ref. \cite{[23]} tells us that ${\cal F}_{1}$ and
${\cal F}_{2}$ are entire functions whose canonical product has genus $1$.
In other words, by virtue of the asymptotic behaviour of the
eigenvalues, one finds that
$$
\sum_{i=1}^{\infty}{1\over |\mu_{i}|}=\infty \; \; {\rm and} \; \;
\sum_{i=1}^{\infty}{1\over |\nu_{i}|}=\infty ,
$$
whereas $\sum_{i=1}^{\infty}{1\over |\mu_{i}|^{2}}$ and
$\sum_{i=1}^{\infty}{1\over |\nu_{i}|^{2}}$ are convergent. This is
why the exponentials $e^{k\over \mu_{i}}$ and $e^{k\over \nu_{i}}$
must appear in Eqs. (A1) and (A2), which are called the
canonical-product representations of ${\cal F}_{1}$ and ${\cal F}_{2}$.
The genus of the canonical product for ${\cal F}_{1}$ is the
minimum integer $h$ such that
$\sum_{i=1}^{\infty}{1\over |\mu_{i}|^{h+1}}$ converges, and
similarly for $F_{2}$, replacing $\mu_{i}$ with $\nu_{i}$.
If the genus is equal to $1$, this ensures that no higher powers of
${k\over \mu_{i}}$ and ${k\over \nu_{i}}$ are needed in the argument
of the exponential. Moreover, even for non-vanishing values of
$\theta$, it remains true that the zeros of ${\cal F}_{1}$ are
minus the zeros of ${\cal F}_{2}$: $\mu_{i}=-\nu_{i}$, for all $i$ [12].
Hence one finds eventually
\begin{equation}
{\cal F}(k)={\widetilde \gamma}k^{2(n+d/2-1)}
\prod_{i=1}^{\infty}\left(1-{k^{2}\over \mu_{i}^{2}}\right),
\label{(A3)}
\end{equation}
where ${\widetilde \gamma} \equiv \gamma_{1}\gamma_{2}$,
$\mu_{i}^{2}$ are the positive zeros of ${\cal F}(k)$, and
the sum $(g_{1}+g_{2})(k)$ can be shown to vanish exactly as
in Sec. IV of Ref. [12].

In our paper we use uniform asymptotic expansions of regular
Bessel functions $J_{\nu}$ and their first derivatives
$J_{\nu}'$. On making the analytic continuation
$x \rightarrow ix$ and then defining
$\alpha_{\nu} \equiv \sqrt{\nu^{2}+x^{2}}$, one can write
\begin{equation}
J_{\nu}(ix) \sim {(ix)^{\nu}\over \sqrt{2\pi}}
\alpha_{\nu}^{-1/2}e^{\alpha_{\nu}}
e^{-\nu \log(\nu+\alpha_{\nu})}\Sigma_{1},
\label{(A4)}
\end{equation}
\begin{equation}
J_{\nu}'(ix) \sim {(ix)^{\nu-1}\over \sqrt{2\pi}}
\alpha_{\nu}^{1/2}e^{\alpha_{\nu}}
e^{-\nu \log(\nu+\alpha_{\nu})}\Sigma_{2},
\label{(A5)}
\end{equation}
where the functions $\Sigma_{1}$ and $\Sigma_{2}$ admit the
asymptotic expansions
$$
\Sigma_{1} \sim
\sum_{k=0}^{\infty}u_{k}(\nu / \alpha_{\nu})/\nu^{k}, \; \;
\Sigma_{2} \sim
\sum_{k=0}^{\infty}v_{k}(\nu / \alpha_{\nu})/\nu^{k} ,
$$
valid uniformly in the order $\nu$ as $|x| \rightarrow \infty$.
The functions $u_{k}$ and $v_{k}$ are polynomials, given by
Eqs. (9.3.9) and (9.3.13) on page 366 of Ref. \cite{[21]}.

\end{document}